\documentclass[structabstract]{aa}  
\usepackage{graphicx}
\usepackage{txfonts}
\usepackage{natbib}
\usepackage{enumerate}
\usepackage{subfig}
\bibpunct{(}{)}{;}{a}{}{,} 
\begin{document}

\title{Multiwavelength study of TeV Blazar Mrk421 during giant flare}
\titlerunning{Multiwavelength study of TeV Blazar Mrk421 during giant flare}

\author{A. Shukla$^{1}$, V. R. Chitnis$^{2}$, P. R. Vishwanath$^{1}$, B. S. Acharya$^{2}$, G. C. Anupama$^{1}$, P. Bhattacharjee$^{3}$, R. J. Britto$^{2,3}$, T. P. Prabhu$^{1}$, L. Saha$^{3}$, B. B. Singh$^{2}$}
\authorrunning{HAGAR collaboration}

\institute{
\inst{1}~Indian Institute of Astrophysics, II Block, Koramangala, Bangalore, 560 034, INDIA.\\ 
\inst{2}~Tata Institute of Fundamental Research, Homi Bhabha Road, Colaba, Mumbai, 400 005, INDIA.\\ 
\inst{3}~Saha Institute of Nuclear Physics, 1/AF, Bidhannagar, Kolkata, 700 064, INDIA.\\ 
\email{amit@iiap.res.in} \\
}
\abstract
%
{ The nearby (z=0.031) TeV blazar Mrk421 was reported to be in a high state of flux activity since November, 2009.}
{ To investigate possible changes in the physical parameters of Mrk421 during its high state of activity using multiwavelength data.}
{We have observed this source in bright state using High Altitude GAmma Ray (HAGAR) telescope array at energies above 250 GeV during February  13\,--\,19, 2010. Optical, X-ray and $\gamma$-ray archival data are also used to obtain the SEDs and light curves.}
%
{ Mrk421 was found to undergo one of its brightest flaring episodes on February 17, 2010 by various observations in X-rays and $\gamma$-rays. HAGAR observations during February 13\,--\,19, 2010 at the energies above 250 GeV show an enhancement in the flux level, with a maximum flux of $\sim$7 Crab units being detected on February 17, 2010. We present the spectral energy distributions during this flaring episode and investigate the correlation of the variability in X-ray and $\gamma$-ray bands.}
{Our multiwavelength study suggests that the flare detected during February 16 and 17, 2010 could arise due to a passing shock in the jet. }


\maketitle

\section{Introduction}
\label{sec:intro}
Blazars are a subclass of Active Galactic Nuclei (AGN) characterized by nonthermal emission extending from radio to high 
energies. The broadband radiation originates within a relativistic jet that is oriented very close to the line of sight. 
Spectral energy distribution (SED) of high energy peaked TeV blazars show two broad peaks. The first peak is located 
between infrared to X-ray energies and the second peak at $\gamma$-ray energies. It is believed that the first peak of the 
SED originates due to synchrotron radiation by relativistic electrons gyrating in the magnetic field of the jet. The 
origin of the high energy GeV/TeV peak is still under debate. This high energy peak might originate either due to interaction of electrons with photon field via Inverse Compton (IC) scattering as in leptonic models or due to interactions of protons with matter, magnetic field or photon fields in hadronic models. The seed photon field for IC scattering in leptonic models may come from
synchrotron emission by the same population of electrons which produce the low energy bump (for a recent review of observations and models, see \cite{2004NewAR..48..367K}) in synchrotron self-Compton (SSC) model and from the external photons from accretion disk \citep{1993ApJ...416..458D} or broad line region (BLR) \citep{1996MNRAS.280...67G} in external Compton models. The hadronic models suggest that high energy peak could be due to extremely energetic protons gyrating in a strong magnetic field emitting synchrotron radiation \citep{2000NewA....5..377A,2003APh....18..593M}, or as inverse Compton and synchrotron emission from a proton-induced cascade \citep{1998Sci...279..684M}. 

The blazar, Mrk421 ($z=0.031$), is the first extragalactic source to be detected at $\gamma$-ray energy $E>500$ GeV
\citep{1992Natur.358..477P}. Since the first detection by the Whipple Observatory $\gamma$-ray telescope in 1992, Mrk421
has been detected by various other Atmospheric Cherenkov Telescopes (ACT) and air shower experiments 
\citep{1996A&A...311L..13P,1997ApJ...490L.141Z,1999A&A...350..757A,2007Ap&SS.309..111B,2007APh....27..447Y,2010A&A...519A..32A,2010MmSAI..81..326D}, and its $\gamma$-ray flux has been found to be highly variable.

Mrk421 was reported to be in a high state of activity during November 2009 to  April 2010, with flaring behavior in X-ray \citep{2010PASJ...62L..55I} and $\gamma$-ray bands detected in February, 2010. One of the brightest flaring episodes of 
this source was observed by various experiments on February 17, 2010. Preliminary results from different experiments show rapid flux variation in the very high energy (VHE) $\gamma$-rays from minute to hour time scales. During this bright 
outburst, the maximum  VHE $\gamma$-ray flux ($>100$ GeV) reached above 9 Crab units \citep{2010ATel.2443....1C}, with 
an average flux of 4 Crab units.

Using the newly commissioned High Altitude GAmma Ray (HAGAR) telescope system, we observed Mrk421 in its high state of 
activity during February to April, 2010 and also detected a very bright flare above 250 GeV. 

In this paper, we study the multiwavelength behavior of Mrk421 during its high state of activity in February, 2010 and 
follow the evolution of its SEDs over a period of seven days, based on data from HAGAR, {\it Fermi}-LAT, RXTE-PCA and 
Swift-XRT. We also present a brief introduction to the HAGAR telescope and the data analysis techniques used to detect 
point like $\gamma$-ray sources with HAGAR.

\label{sec:observation}
\section{HAGAR}

HAGAR, an array of ACTs using wave front sampling technique, is located at the Indian Astronomical Observatory (IAO), 
Hanle (32$^\circ$ 46$^{\prime}$ 46$^{\prime\prime}$ N, 78$^\circ$ 58$^{\prime}$ 35$^{\prime\prime}$ E), in the Ladakh region of
India, at an altitude of 4270 m. The main motivation behind setting up the $\gamma$-ray array at a high altitude is to 
exploit the higher Cherenkov photon density and thus achieve lower energy threshold \citep{2001ICRC....7.2769C}. HAGAR 
consists of an array of seven telescopes in the form of a hexagon, with one telescope at the center. All the seven 
telescopes have seven para-axially mounted front coated parabolic mirrors of diameter $0.9$ m, with a UV sensitive 
photo-tube at the focus of individual mirrors. Each telescope is separated by $50$ m distance from its neighboring 
telescope. The seven PMT pulses of a telescope are linearly added to form a telescope output called the {\it Royal sum} 
pulse. A coincidence of at least 4 {\it Royal sum} pulses out of seven, above a predetermined threshold, is taken within 
a time window of 150 ns or 300 ns depending on the zenith angle of pointing direction to generate a trigger for 
initiating data recording. The HAGAR Data Acquisition (DAQ) system is CAMAC based. Data recorded for each event consist 
of
\begin{enumerate}[a.] 
\item Relative arrival time of the Cherenkov shower front at each mirror, as measured by TDCs with a resolution of 250 ps.
\item A Real Time Clock (RTC) module synchronized with GPS is used to record the absolute arrival time of these events 
accurate upto $\mu$s. 
\item The density of Cherenkov photons at each telescope is measured by the total charge present in PMT pulses (this is 
recorded by using 12 bit QDCs). 
\item Information on {\it Royal sum} pulses are also recorded like the individual PMT pulses.
\item Latch information to indicate the triggered telescopes and other house keeping information on various scalar 
readings.
\end{enumerate}

In addition, a parallel DAQ using commercial waveform digitizers with a sampling rate of 1 GS/s (ACQIRIS make model DC271A) is also used. The direction and energy of $\gamma$-rays are estimated by measuring relative time delays and densities of 
Cherenkov photons at each telescope respectively.

The performance of the HAGAR array has been studied by simulations, which are done in two steps: (1) Cherenkov emission due to $\gamma$-ray and cosmic ray induced air showers
in the atmosphere, by using the Monte Carlo simulation package CORSIKA, 
developed by the KASCADE group \citep{1998cmcc.book.....H}, (2) study of the response of the array towards the Cherenkov 
radiation produced by the simulated showers. The performance parameters such as energy threshold, collection area and 
sensitivity of the experiment are obtained by a detector simulation package indigenously developed by the HAGAR 
collaboration. Energy threshold of the HAGAR telescope is estimated as 204 GeV in case of vertically incident 
$\gamma$-ray showers for a $\ge$ 4-Fold trigger condition, for which the corresponding collection area is 
$3.2\times10^{8}$ cm$^{2}$. HAGAR sensitivity is such that it will detect a Crab nebula like source at a significance 
level of $5\sigma$ in 15 hours of observation \citep{2011ICRC....OG2.5}.

Cherenkov emission due to induced air showers forms a spherical wavefront with a large radius of curvature and thickness 
of $\sim$ 1 m at the observation level. This Cherenkov emission mainly originates at the shower maximum region which is 
at a height of about 5 km above the ground level at Hanle. This spherical wavefront is approximated as a plane wavefront in the data analysis procedure, which is a good approximation at the observation level. The arrival direction of each 
shower in Cherenkov light pool is computed by measuring the relative arrival times of shower front at different 
telescopes. The normal to this plane front gives the arrival direction of the incident shower. The angle between the
direction of the shower axis and the pointing direction of the telescope is defined as Space Angle ($\psi$). The Space 
Angle is estimated for every event by measuring the relative arrival time of the shower front at each telescope.

The observations were carried out by pointing all seven telescopes towards the source or background direction at a time. 
Each source run was followed (or preceded) by a background run with the same exposure time (typically $40$ minutes) and 
covered the same zenith angle range as that of the source to ensure that observations were carried out at almost the 
same energy threshold. Data selection was done by using parameters which characterize good quality data, in order to 
reduce systematic errors.

\begin{figure}
\centering
\includegraphics[width=9.1cm]{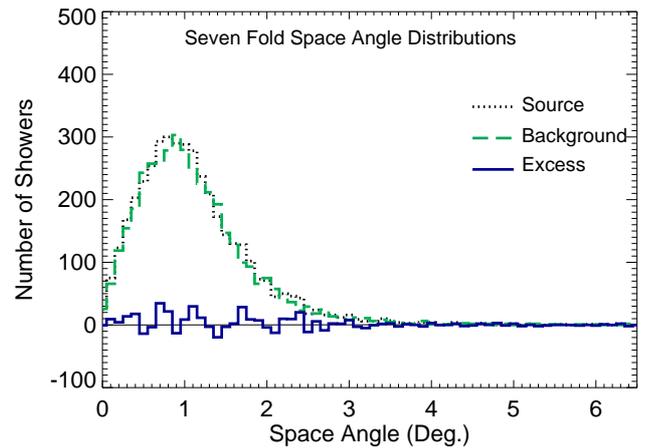}
\caption{Typical Space Angle Distribution plot obtained by HAGAR for a seven fold trigger.}
\label{fig:spang}
\end{figure}

Extraction of the $\gamma$-ray signal was carried out by comparing the ON source and OFF source space angle 
distributions obtained during the same night. The excess events were computed from 0$^\circ$ to Lower Limit (LL), where 
LL is defined as the foot of distant half maximum point computed by fitting a Gaussian function to the space angle 
distribution. Background space angle distribution was normalized with source space angle distribution by comparing the 
tails of the distributions (LL to 6.5$^\circ$), since no $\gamma$-ray events are expected in this region. This 
normalization is required to compensate for possible changes in observation conditions or sky conditions during ON 
source and OFF source regions. Thus the number of $\gamma$-ray events are estimated as 
\begin{equation}
\textrm{Number of}~\gamma-\textrm{rays}=\Sigma_{0}^{LL}(S_{i}-c_{k}B_{i}),
\end{equation}
where $S_{i}$ and $B_{i}$ are source and background events respectively and $c_{k}$ is the normalization constant 
obtained using the tail of the space angle distribution.

A typical space angle plot obtained for seven fold trigger is shown in Figure \ref{fig:spang}. The dotted (black) 
line histogram corresponds to source observations, dashed (green) line histogram corresponds to background 
observations. The histogram shown as solid (blue) line corresponds to the estimation of $\gamma$-ray events computed as 
the excess of ON source events over the normalized OFF source background events.

\section{Multiwavelength observations and Analysis} 
Data obtained on Mrk421 by HAGAR during its high state in 2010 are combined with archival data from 
{\it Fermi}-LAT, RXTE-ASM, Swift-BAT, Swift-XRT, RXTE-PCA, SPOL and OVRO for a multiwavelength study. Details of the observations and 
analysis procedure used for each data set are discussed in this section.

\subsection{HAGAR}
The HAGAR observations of Mrk421 were carried out for three months on moonless nights, during February to April, 2010 at
a mean zenith angle of 6$^\circ$. A total of 30 hours of data were collected during these observations. Suitable cuts were 
applied to the difference in the trigger rates of ON source and OFF source regions to ensure that observations were 
carried out at almost the same energy threshold. Some data were also rejected due to unavailability of all seven 
telescopes at the time of observations. After imposing such data quality cuts, a total of 18 hours of observation is 
used for further analysis. Data were analyzed according to the procedure discussed in \S 2 and 
\cite{2007Ap&SS.309..111B}. We have analyzed only events with signals in at least 5 telescopes ($\ge$ 5-Fold) in order 
to reduce systematic errors in our data. The $\ge$ 5-Fold events correspond to an energy threshold of about 250 GeV. 

\subsection{{\it Fermi}-LAT} 

The Large Area Telescope (LAT) is a pair production telescope \citep{2009ApJ...697.1071A} on board {\it Fermi} spacecraft. It 
covers the energy range from 20 MeV to more than 300 GeV with a field of view $\ge$ 2.5 sr. {\it Fermi}-LAT data\footnote{http://fermi.gsfc.nasa.gov/} of Mrk421 obtained during 2010 February 12\,--\,22 above 100 MeV were analyzed using the standard
analysis procedure {\it(ScienceTools)} provided by the {\it Fermi}-LAT collaboration. A circular region of 10$^\circ$
 radius ``region of interest (ROI)'' was chosen around Mrk421 for event reconstruction from the so-called ``diffuse'' event 
class data which has the maximum probability of being the source photons. We only retained events having a zenith angle 
$<$105$^\circ$ to avoid the background from Earth albedo. The spectral analysis on the resulting data set was carried out by 
including galactic diffuse emission component model (gll$\_$iem$\_$v02.fit) and an isotropic background component model 
(isotropic$\_$iem$\_$v02) with post-launch instrumental response function P6V3 DIFFUSE, by using unbinned maximum likelihood 
analysis \citep{1979ApJ...228..939C,1996ApJ...461..396M}. A power law spectrum was used to model the source spectrum 
above 100 MeV, with integral flux and photon index as free parameters. The flux, spectrum and source location are 
determined by using unbinned GTLIKE algorithm.

\subsection{X-ray data from RXTE and Swift} 

Proportional Counters Array (PCA) \citep{1993A&AS...97..355B} is an array of five identical xenon filled proportional 
counter units (PCUs). The PCUs cover energy range from 2\,--\,60 keV with a total collecting area of 6500 cm$^{2}$. The 
archival X-ray data from PCA on board RXTE during 2010 February 13\,--\,19 was analyzed to obtain the X-ray spectrum and 
light curve. We have analyzed standard 2 PCA data which has a time resolution of 16 seconds with energy information in 
128 channels. Data analysis was performed using FTOOLS (version 5.3.1) distributed as a part of HEASOFT (version 5.3). 
Data were filtered using the standard procedure given in the RXTE Cook Book 4 for each of the observations. The 
background models were generated with the tool ``pcabackest'', based on RXTE GOF calibration files for a 'bright' source 
(more than 40 ct/sec/PCU). The PCA spectrum in the the energy range of 3\,--\,30 keV was fitted by using XSPEC with a 
cutoff powerlaw with line of sight absorption. The line of sight absorption was fixed to neutral hydrogen column density 
at  1.38$\times$10$^{20}$ cm$^{-2}$ \citep{1990ARA&A..28..215D}. 

The XRT on board Swift \citep{2005SSRv..120..165B} uses a grazing incidence Wolter I telescope to focus X-rays onto a CCD. The instrument has an effective area of 110 cm$^{2}$, 23.6 arcmin FOV, 15 arcsec resolution (half-power diameter), and an energy range of 0.2\,--\,10 keV. The Windowed Timing (WT) mode data were used to obtain spectrum (0.3\,--\,3 keV) from Swift-XRT during 2010 February 13\--\,19. Source photons were extracted using a box region with the length of 40 pixels and width about 20 pixels. Events with grades  0\,--\,2 were selected for the WT mode data. The spectral data were rebinned by GRPPHA 3.0.0 with 20 photons per bin. Standard auxiliary response files and response matrices were used. Spectra for
this source were fitted using XSPEC version 12.3.1 with a model consisting of absorbed power law over the energy range 
of 0.3\,--\,10 keV.

The ``Dwell'' data from RXTE-ASM was obtained from ASM website\footnote{http://xte.mit.edu/} and these data were analyzed
by the method discussed in \cite{2009ApJ...698.1207C}. A daily average flux between 15\,--\,50 keV from Swift-BAT was 
obtained from BAT website\footnote{http://heasarc.nasa.gov/docs/swift/results/transients/}.

\subsection{Optical and radio data}
 
The optical and radio data made available in the {\it Fermi} multiwavelength support program websites\footnote{http://james.as.arizona.edu/~psmith/Fermi/}\fnmsep\footnote{http://www.astro.caltech.edu/ovroblazars/index.php?page=home} are used in this study. 

The optical observations were made using the SPOL CCD Imaging/Spectropolarimeter at Steward Observatory \citep{2009arXiv0912.3621S}. The optical V band photometric data were obtained.

The 15 Ghz radio observations were made by using a 40 meter single dish telescope at Owens Valley Radio Observatory (OVRO). Details of the analysis are described in \cite{2011ApJS..194...29R}. 

\section{Results} 

 Mrk421 was found to be in a high state of activity during the entire period of HAGAR observations during February\,--\,April, 2010,
and was in its brightest state in February, 2010. The $\gamma$-ray and X-ray fluxes decreased in the later months, but were still higher than that during the quiescent state. Results of observations of Mrk421 using HAGAR during the three months period are given in Table 1.

\begin{figure}
\centering
\includegraphics[width=8.3cm]{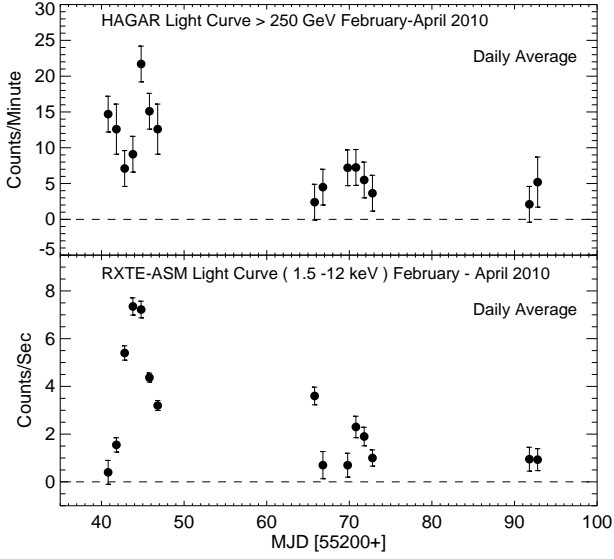}
\caption{The top panel shows the daily average light curve of Mrk421 during the period February\,--\,April, 2010 in VHE 
$\gamma$-rays above 250 GeV from HAGAR and the bottom panel shows the X-ray light curve in 1.5\,--\,12 kev from RXTE-ASM, during 
the same period.}
\label{fig:image0}
\end{figure}

 Figure \ref{fig:image0} contains the daily light curve of Mrk421 during February\,--\,April, 2010. The upper panel shows the daily average of VHE $\gamma$-ray flux obtained from HAGAR. Bottom panel shows daily average from ASM on board RXTE in 1.5\,--\,12 keV. It is clearly seen in the HAGAR as well as RXTE\,--\,ASM light curves that Mrk421
was in its brightest state in February, 2010 in both VHE $\gamma$-rays and X-rays.

HAGAR telescope detected Mrk421 in a high state of VHE $\gamma$-ray flux, during the period February 13\,--\,19, 2010.
One of the brightest flaring episodes was observed on February 17, 2010 (see Figure \ref{fig:image0}), with the maximum 
flux between 6\,--\,7 Crab units. The source was detected with 5 sigma significance in less than 40 minutes of 
observations. We investigate in the following the multiwavelength behaviour of Mrk421 during the flaring episode.

\begin{table}
\small
\caption{HAGAR observations during high state of activity.}
\begin{tabular}{ccccc}
\hline
\hline
Month & Total & Excess number & Mean & Signi- \\
2010& duration & of on source & $\gamma$-ray rate& ficance \\
&   (min)       & events       & (/min) & $\sigma$\\
\hline
February   & 479       &6418.22  & 13.4$\pm$1.05  &12.7  \\
March      & 478       &2524.23  & 5.3$\pm$1.1  & 4.8  \\
April      &116        &414      & 3.5$\pm$2.1  & 1.7 \\
\hline
\end{tabular}
\label{tab:red_oscil}
\end{table}

\begin{figure}
\centering
\includegraphics[width=10cm]{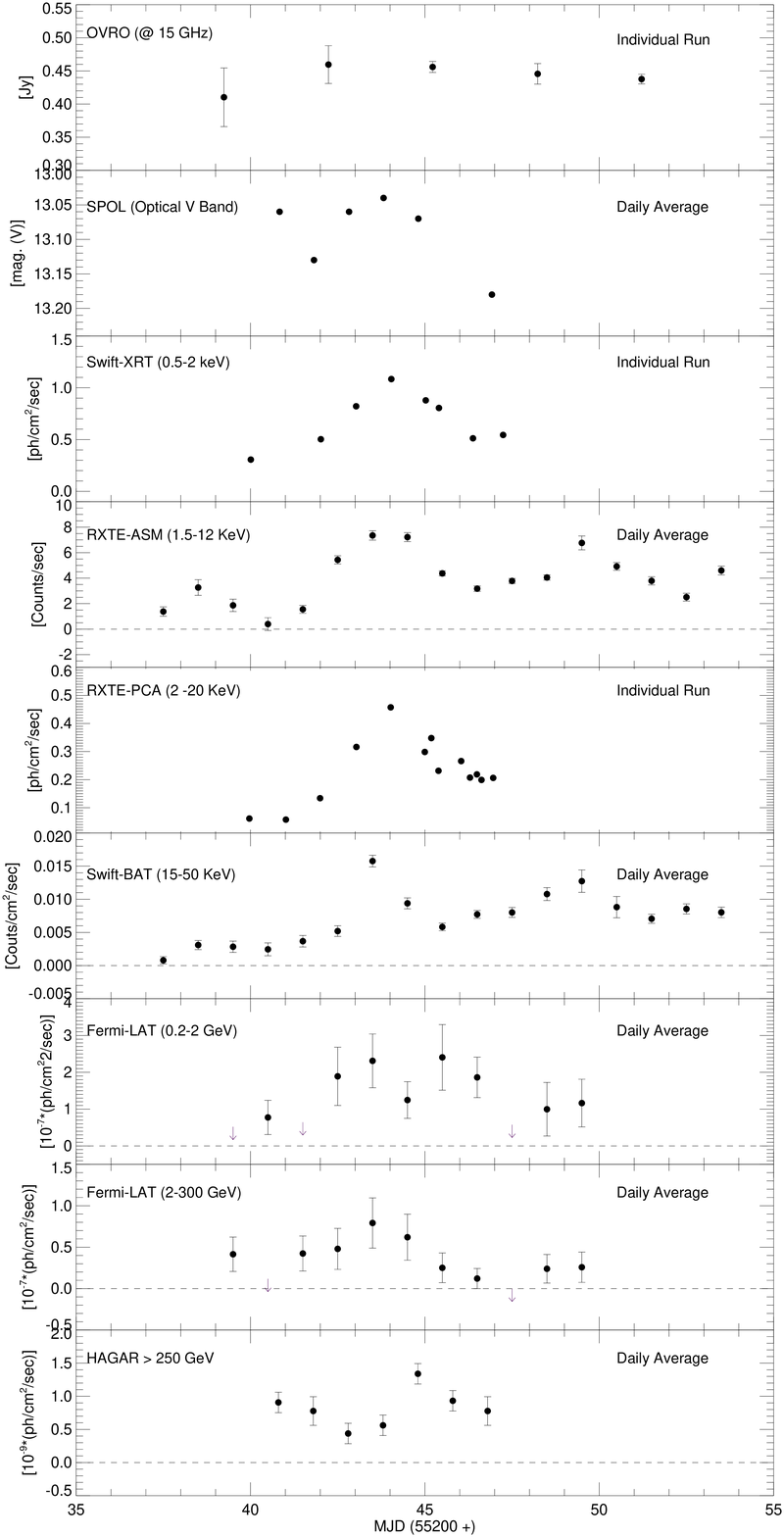}
\caption{ Multiwavelength light curve of Mrk421 during February 10\,--\,26, 2010.
\label{fig:image2}
}
\end{figure}

\subsection{Flux Variability during February 10\,--\,26, 2010}

The multiwavelength (radio to $\gamma$-rays) quasi-simultaneous light curve of Mrk421 during February 10\,--\,26, 2010 based on observations in \S 3, is shown in Figure \ref{fig:image2}.
The top eight panels correspond to data from OVRO, SPOL, Swift\,--\,XRT, RXTE\,--\,ASM, RXTE\,--\,PCA, Swift\,--\,BAT, {\it Fermi}-LAT (0.2\,--\,2 GeV) and {\it Fermi}-LAT (2\,--\,300 GeV) 
respectively. The bottom panel corresponds to HAGAR data above 250 GeV.

A clear variation of flux over a period of seven days is observed in the optical, X-rays and $\gamma$-rays during February
13\,--\,19, 2010. The peak flux in optical, X-rays and low energy $\gamma$-rays are observed to be around February 16, 2010 while in 
the VHE $\gamma$-ray band, the peak is seen on February 17, 2010 with a possible one day lag 
compared to the situation at lower energies. The radio flux has not changed significantly during this time.

\begin{figure}
\centering
\includegraphics[width=8cm]{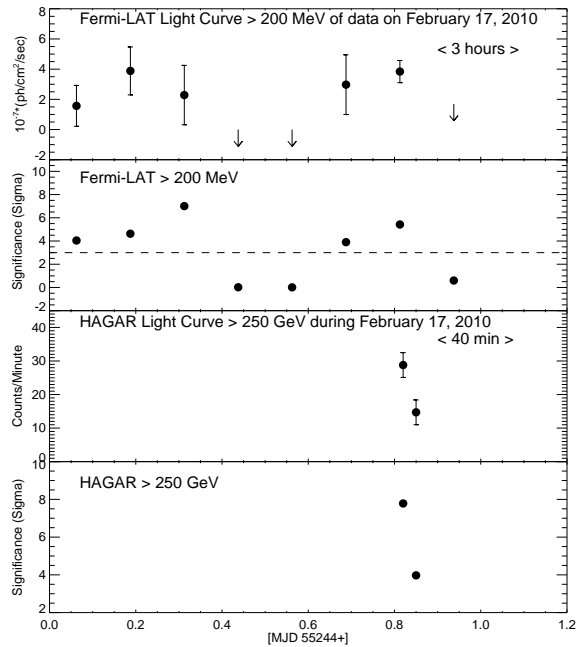}
\caption{ {\it Fermi}-LAT and HAGAR intra-day light curve of Mrk421 during February 17, 2010.
\label{fig:image11}
}
\end{figure}

\subsection{Intra-Day and spectral variability during February 10\,--\,26, 2010}

 {\it Fermi}-LAT data indicate an intra-day flux variability at energies $>$ 200 MeV during the TeV flare on February 17, 2010. An increase in the flux is seen in the first 9 hours (MJD55244.0\,--\,55244.4) of LAT observations, followed by a quiescence that lasts for a few hours (MJD55244.4\,--\,55244.6) before the occurrence of another increase in flux during MJD55244.7 to MJD55244.85. A similar trend was detected by the VERITAS collaboration 
\citep{2010ATel.2443....1C}. HAGAR also observed a continuous decrease in the flux over a period of $\sim$ 2 hrs (MJD55244.8\,--\,55244.86), which were simultaneous with LAT during later part of the night (see Figure 4).

\begin{figure}
\centering
\includegraphics[width=9.1cm]{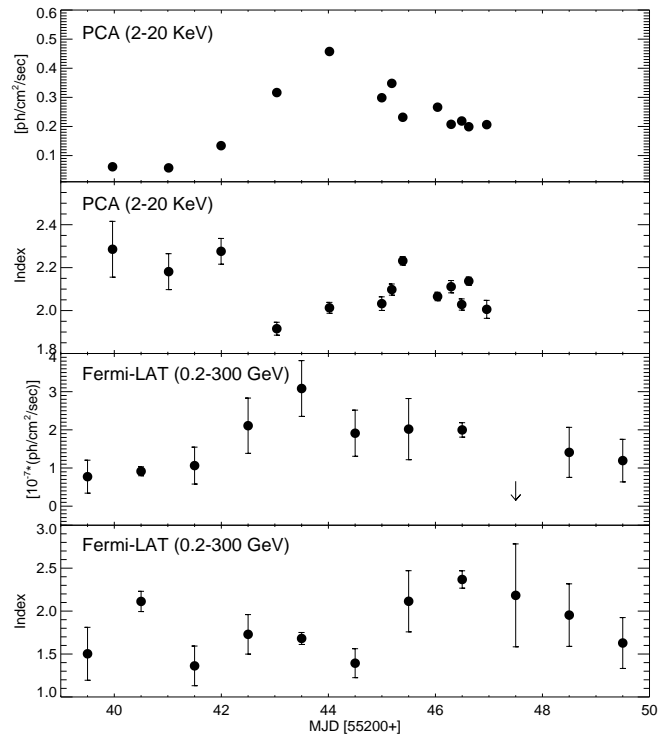}
\caption{RXTE\,--\,PCA light curve and spectral index are plotted in the first two panels, {\it Fermi}-LAT light curve in the range
0.2\,--\,300 GeV and the photon index are plotted in bottom two panels.
\label{fig:image3}
}
\end{figure}

Spectral variation also has been detected over eleven days (February 12\,--\,22) of {\it Fermi}-LAT observations and seven days 
(February 13\,--\,19) of RXTE-PCA observations. A photon index of $1.39 \pm 0.17$, implying a flat/hard energy spectrum 
was observed in the {\it Fermi}-LAT data at energies above 100 MeV on February 17, 2010. On the other hand maximum hardening in RXTE-PCA data was observed on 
February 16 (Figure \ref{fig:image3} and Table 2). Spectral hardening in the X-rays \citep{2004ApJ...601..759T} and $\gamma$-rays \citep{1997ApJ...490L.141Z} during strong flares has been reported earlier also.
\begin{table}
\small
\caption{Cutoff energy (PCA), spectral index (PCA) and photon index ({\it Fermi}-LAT $>$100 MeV) for four different activity 
states during February 13\,--\,19.}
\begin{tabular}{cccc}
\hline\hline
\textbf{State}&\textbf{Cutoff PCA (keV)} & \textbf{Index PCA } & \textbf{Index LAT} \\
\hline
Pre Flare      & 29.1$\pm$6.3     &2.34$\pm$0.05  &1.65$\pm$0.03\\
Moderate Flare & 19.5$\pm$1.5     &1.91$\pm$0.03  &1.67$\pm$0.15\\
TeV Flare      & 26.8$\pm$2.2     &2.03$\pm$0.02  &1.39$\pm$0.16\\
Post Flare     & 35.2$\pm$2.3     &2.19$\pm$0.02  &2.06$\pm$0.04\\
\hline
\end{tabular}
\label{tab:red_oscil}
\end{table}

\section{Discussion}

Mrk421 has shown several high states of flux activity in X-rays and $\gamma $-rays over the period of the past two 
decades \citep{1996ApJ...472L...9B,1996Natur.383..319G,1999ApJ...511..149K,2000ApJ...542L.105T,2002BASI...30..385B,
2005ApJ...630..130B,2007Ap&SS.309..111B,2007A&A...462...29G,2007ApJ...663..125A,2008ApJ...677..906F,2009ApJ...695..596H,
2011ASTRA...7...65V,2010APS..4CF.L1004B}. It was observed in the brightest state of TeV $\gamma$-ray flux  during the 
years of 2000-2001, when its maximum flux reached as high as 14 Crab units above 1 TeV \citep{2010A&A...524A..48T}. It 
has also shown several mild high states since 2001, but was not found to be in an extreme high state till 2010. Mrk421 
had brightened up again in November 2009 and was observed by the VERITAS collaboration in an extreme high flux state on 
February 17, 2010. The VERITAS collaboration reported a maximum flux of 9 Crab units above 100 GeV \citep{2010ATel.2443....1C} 
and the HESS collaboration \citep{2011arXiv1106.1035T} also found this source in a high state during their follow up 
observations. 

 We have also detected this flare using the HAGAR telescope system, a few hours after the VERITAS observations, with a 
maximum flux of 6\,--\,7 Crab units above 250 GeV (1 Crab unit $\sim$ 4.2 $\gamma$-rays/minute above 250 GeV). HAGAR 
continued the observations of Mrk421 over the next two months. The average flux during February, 2010 was found to be 
$\sim$ 3 Crab units and 1 Crab unit during March and April, 2010.

\subsection{Spectral Energy Distribution}

 The Spectral Energy Distribution of TeV blazars has a typical two-bump shape. It is often believed that the broadband 
emission from these sources is produced by synchrotron self-Compton (SSC) mechanism. The lower energy bump, which peaks 
at infrared to X-ray is interpreted as synchrotron emission emitted by relativistic electrons gyrating in the magnetic 
field of the jet, and the GeV/TeV bump is attributed to IC scattering of synchrotron photons by the same population of 
electrons which produces the synchrotron radiation.

A one zone homogeneous SSC model developed by \cite{2004ApJ...601..151K} is used to fit the multiwavelength data to 
obtain the SED. This model assumes a spherical blob of plasma of a comoving radius $R$ which travels with a bulk 
Lorentz factor $\Gamma$ towards the observer. The emission volume is filled with an isotropic population of electrons and 
a randomly oriented uniform magnetic field $B$. Energy spectrum of the injected electrons in the jet frame is described 
by a broken power law with low-energy $(E_{\rm{min}}$ to $E_{\rm{b}})$ and high-energy $(E_{\rm{b}}$ to $E_{\rm{max}})$ 
components with indices of $p1$ and $p2$. The emitted radiation is Doppler boosted by the Doppler factor 
\begin{equation}
\delta=[\Gamma(1-\beta\cos(\theta))]^{-1}
\end{equation}
where $\beta$ is the bulk velocity of the plasma in units of the speed of light and $\theta$ the angle between
jet axis and the line of sight in the observer frame.  The VHE $\gamma$-ray spectrum is corrected for absorption by the 
extragalactic background light \citep{2008A&A...487..837F}.

The radius of the emission zone is constrained by the variability time scales. Variability present in X-ray and 
$\gamma$-ray data is of the order of one day, so we have chosen $t_{VAR}\sim 1$ day. The comoving radius of the emission 
zone is defined as 
\begin{equation}
 R\sim{c\delta t_{var}}/{(1+z)}
\end{equation}

 We have attempted to obtain SEDs for different flux states using multiwavelength data of the Mrk421 during HAGAR 
observations of February 13\,--\,19, 2010 in an attempt to search for any changes in the physical parameters over this 
period. This multiwavelength data is divided into four states according to the flux state of the source as follows: 
State 1: Pre Flare (13\,--\,15 February), State 2: moderate flare (16 February), State 3: TeV flare (17 February) 
and State 4: post flare (18\,--\,19 February). The pre flare state of the source is a moderate high state. 

The {\it Fermi}-LAT data is divided into three bins (0.1\,--\,1 GeV, 1\,--\,3 GeV and 3\,--\,300 GeV) to obtain spectrum of 
Mrk421 for 'State 1', 'State 2' and 'State 4' by freezing the photon index to 1.65, 1.67 and 2.06 respectively. These photon 
indices were obtained by analyzing the 0.1\,--\,300 GeV data from {\it Fermi}-LAT. The {\it Fermi}-LAT spectrum of February 17, 2010 was 
obtained by dividing {\it Fermi}-LAT data into four bins (0.1\,--\,1 GeV, 1\,--\,3 GeV, 3\,--\,10 GeV and 10\,--\,300 GeV) by 
freezing the photon index to 1.39, obtained by the analysis of the 0.1\,--\,300 GeV data. The best fit photon indices and cutoff energies for all four states are presented in Table 2.

The magnetic field, Doppler factor ($\Gamma\sim\delta$), electron energy density, break present in electron injection spectrum $(E_{\rm{b}})$, $p1$, $p2$, minimum and maximum 
electron energies are used as free parameters while fitting the model to optical, X-ray and $\gamma$-ray data. The angle between the jet axis and the line of sight in the observer frame
is taken to be 2.6$^\circ$.
\begin{table*}
\small
\begin{center}
\caption{Correlation coefficient and time lag of Mrk421 emission during February\,--\,April, 2010}
\begin{tabular}{cccccc}

\hline\hline
{\textbf{Instruments}} & \textbf{Lag (Days)} & \textbf{Correlation} & \textbf{Algorithm}  & \textbf{Data Used} \\
                       &                      &\textbf{coefficient}  &                     &  \textbf{2010}   & \\
\hline
RXTE\,--\,ASM vs Swift\,--\,BAT             &  0.0                  & 0.81                 &   CCF               &  13 -- 19 February \\   
RXTE\,--\,ASM vs Swift\,--\,BAT             &  0.0                  & 0.77                 &   ZDCF              &  10 -- 27 February \\
RXTE\,--\,ASM vs HAGAR                      &  1.3                  & 0.62                 &   ZDCF              &  February to April \\  
Swift\,--\,BAT vs HAGAR                     &  1.3                  & 0.88                 &   CCF               &  13 -- 19 February   \\  
Swift\,--\,BAT vs HAGAR                     &  1.3                  & 0.82                 &   ZDCF              &  13 -- 19 February   \\   
Swift\,--\,BAT vs HAGAR                     &  1.3                  & 0.74                 &   ZDCF              &  February to April \\  

\hline
\end{tabular}
\label{tab:red_oscil}
\end{center}
\end{table*}

\subsection{Evolution of the SED during the high state}

\begin{figure*}
\centering
\subfloat[]{
 \includegraphics[width=7.1cm]{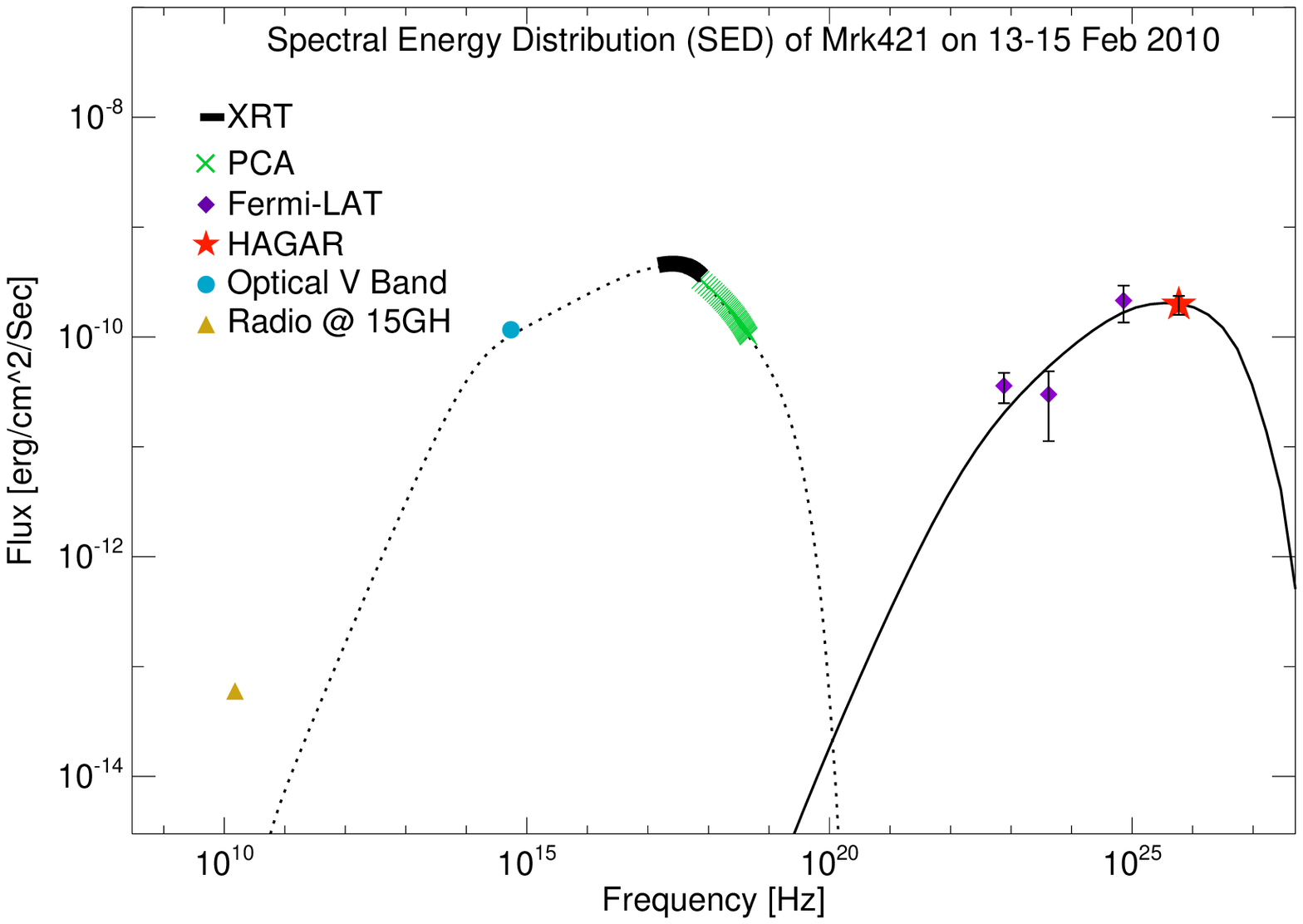}}
\subfloat[]{
 \includegraphics[width=7.1cm]{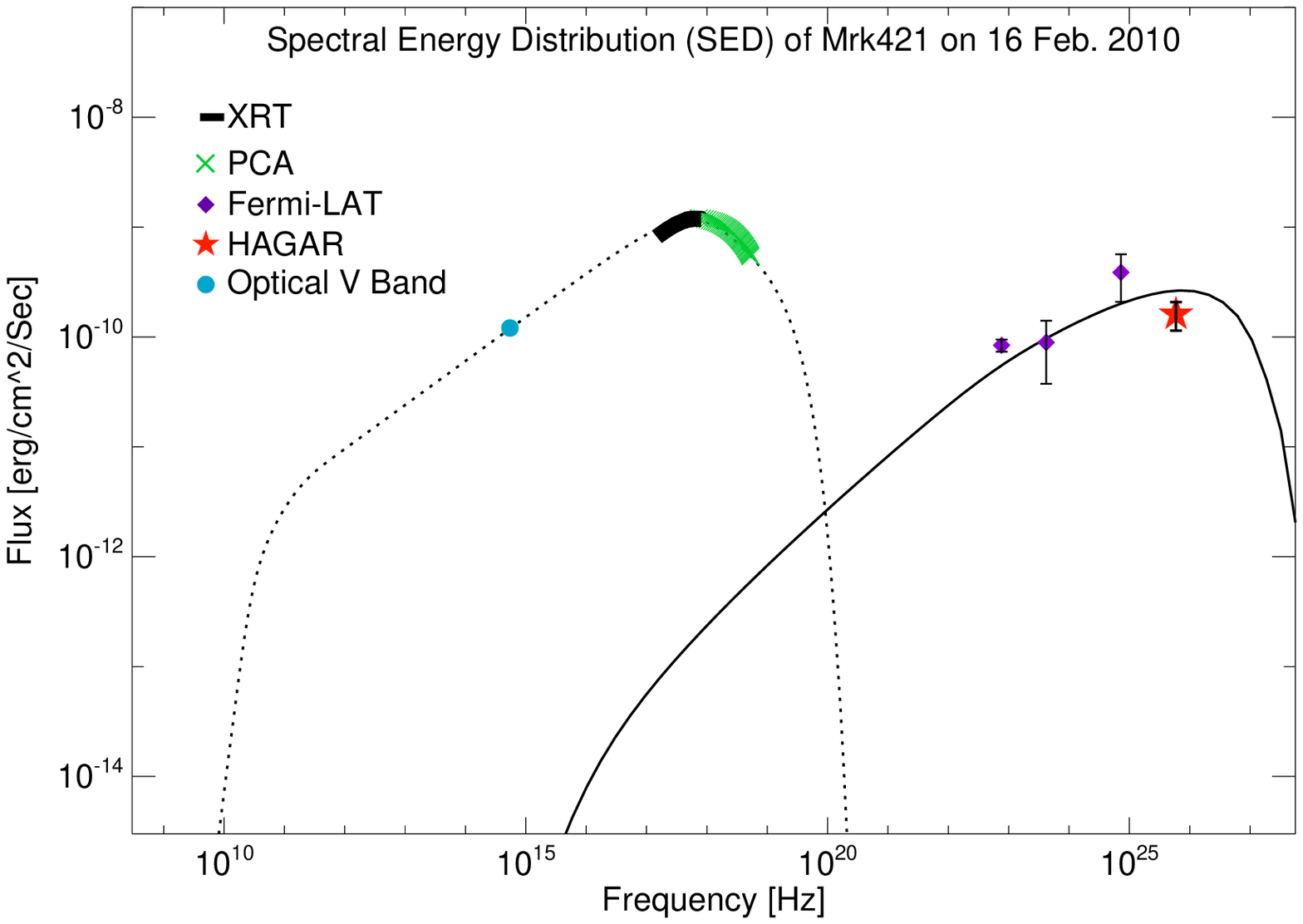}} \\
\subfloat[]{
 \includegraphics[width=7.1cm]{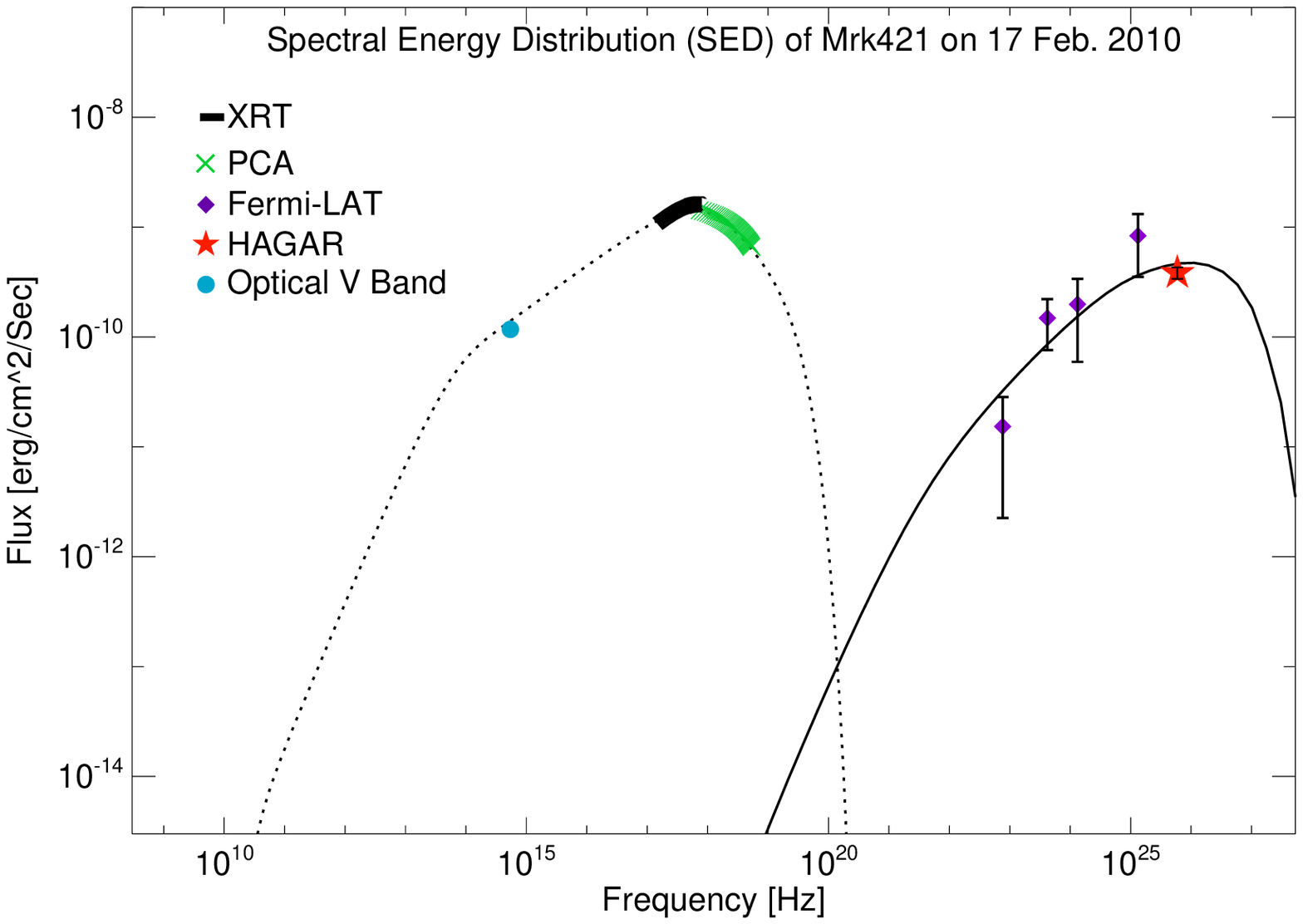}}
\subfloat[]{
\includegraphics[width=7.1cm]{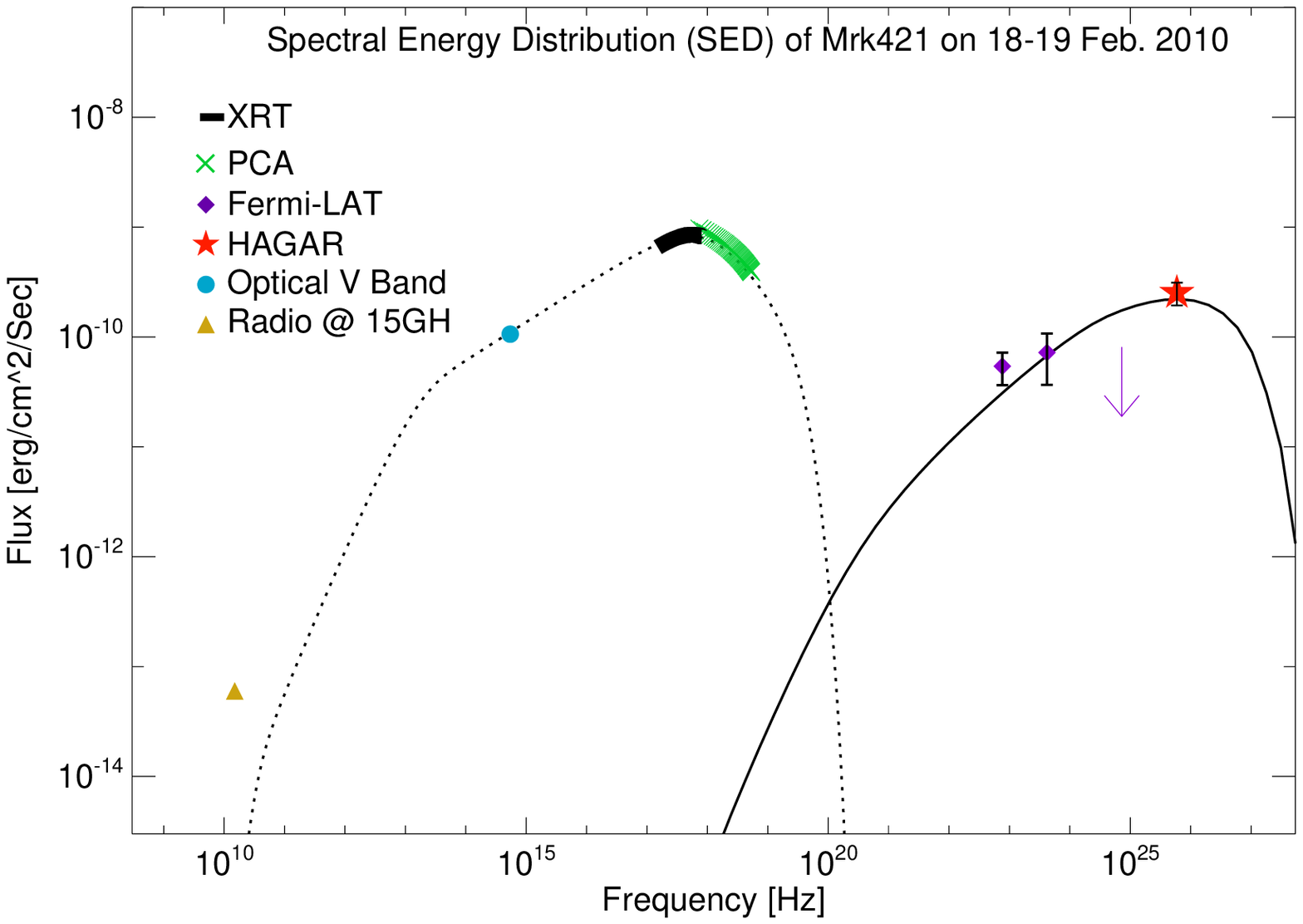}}\\
\centering
\subfloat[]{
\includegraphics[width=9.1cm]{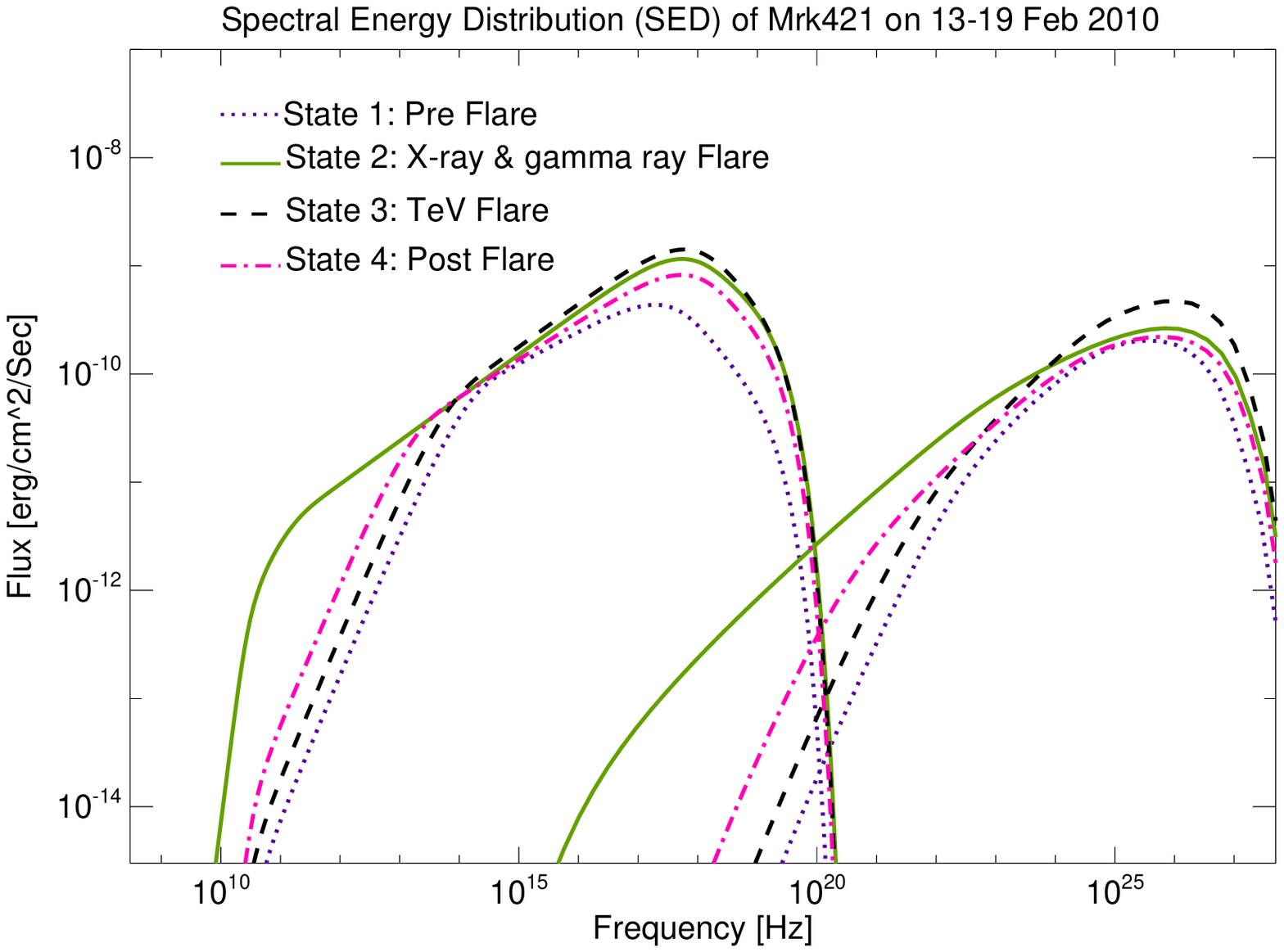}}\\
\caption{SEDs of Mrk421 during February 2010.
\label{fig:image4}
}
\end{figure*}

The {\it Fermi}-LAT collaboration \citep{2009ApJ...707.1310A} had reported photon index of the energy spectrum of Mrk421 to be 
1.78 from the first 5.5 months of their observations. This refers to an average spectrum mostly during the quiescent state. 
However, during 'State 1' (see Figure 6a), we find the energy spectrum to be flatter with a photon index of 1.65 
corresponding to 0.1\,--\,300 GeV {\it Fermi}-LAT data. Also, we see a plateau in the light curves during this state in the 
{\it Fermi}-LAT data (in 2\,--\,300 GeV energy bands) and 15\,--\,50 keV Swift\,--\,BAT data (see Figure 3). The presence of a plateau in the light curve indicates that the injection time scales of energetic electrons are longer than 
the cooling time scales, and the entire volume of emission zone is radiating. Plateaux are also observed in other 
$\gamma$-ray sources like 3c454.3 \citep{2011ApJ...733L..26A}. 

Flares in optical, X-ray, low energy ($<$2 GeV) $\gamma$-ray and HE (2\,--\,300 GeV) $\gamma$-ray band were seen in 'State 2'. The corresponding SED is shown in Figure 6b. These low and HE $\gamma$-ray flares could be caused by upscattering of X-ray photons by low energy electrons which are produced by SSC cooling in 'State 1'. As an effect of low energy flare, the low energy part of SED in 'State 2' is much flatter than that in 'State 1'.

The observed GeV/TeV flares above 250 GeV on 17 February could be produced by upscattering of X-ray photons in Swift\,--\,BAT energy 
range by higher energy electrons (see Figure 6.c for the SED). The presence of a passing shock might have accelerated 
the electrons to higher energies in the emission region.

The source is brighter at lower energy $\gamma$-ray in 'State 4', which could be due to cooling of electrons after the 
high energy flare. The corresponding SED is shown in the Figure 6d.

The derived SED parameters for each 
state are listed in Table 4, and in Figure 6e we illustrate the change that the SED undergoes during the four different 
states of the source.
\begin{table*}
\small
\begin{center}
\caption{SED parameters obtained by fitting to data using t$_{var}$$\sim$ 1 day}
\begin{tabular}{cccccccccccc}
\hline\hline
State &  Magnetic & Doppler & log E$_{min}$\tablefootmark{a} & log E$_{max}$\tablefootmark{b} & E$_{break}$\tablefootmark{c} & p1 & p2 & Sy$_{pk}$\tablefootmark{d} & IC$_{pk}$\tablefootmark{e} & U$_{e}$\tablefootmark{f} & $\eta $\tablefootmark{g} \\ 
-& field & factor & [eV] & [eV] & [eV]& & - & [$10^{17}$] & [$10^{25}$] & [$10^{-3}$] &   $[u_{e}^{'}/u_{B}^{'}]$\\
-& (G) & ($\delta$) & - & - & - & -& & (Hz) & (Hz) & (erg/cc) & -  \\
\hline
State1   & 0.026  & 19.5  & 9.6  & 12.1  & 11.3   & 2.4  & 4.3   &  1.93   &  3.58    & 0.9   &  33.46 \\
State2   & 0.029  & 22.0  & 8.0  & 12.1  & 11.40  & 2.2  & 3.9   &  5.74   &  7.02    & 1.4   &  41.83 \\
State3   & 0.029  & 21.0  & 9.4  & 12.1  & 11.45  & 2.2  & 4.1   &  6.13   &  11.53   & 1.0   &  29.88 \\
State4   & 0.028  & 21.0  & 9.1  & 12.1  & 11.45  & 2.3  & 4.1   &  5.53   &  6.76    & 8.5   &  27.24 \\
\hline
\end{tabular}
\label{tab:red_oscil}
\end{center}
\tablefoottext{a}{E$_{min}$: Minimum value of energy of the electrons present in the emission zone} \\
\tablefoottext{b}{E$_{max}$:Maximum value of energy of the electron present in the emission zone} \\
\tablefoottext{c}{E$_{break}$: Break in the electron injection spectrum} \\
\tablefoottext{d}{Sy$_{pk}$:Peak value of synchrotron bump}\\
\tablefoottext{e}{IC$_{pk}$:Peak value of IC bump}\\
\tablefoottext{f}{U$_{e}$:Electron energy density}\\
\tablefoottext{g}{$\eta $:Equipartition coefficient }
\end{table*}
\subsection{Light Curve}

Significant flux variation in optical, X-ray and $\gamma$-ray bands is detected during February, 2010. The highest energy tail of 
the electron energy distribution ($\gamma \geq \gamma_{br}$) is responsible for the production of the observed X-ray 
synchrotron continuum at $\ge$ 0.5 keV, while the TeV $\gamma$-rays might be produced through upscattering of 
synchrotron photons by the same population of electrons. The observed optical, X-ray variability during February, 2010 may be explained by injection of fresh electron in emission zone and cooling of the electrons due to SSC mechanism. Swift\,--\,BAT light curve showed faster variability than RXTE\,--\,ASM, which could be due to the cooling effect of high 
energy electrons, which produce X-rays at the 15\,--\,50 keV range.

The observed $\gamma$-ray variability mainly divided into two bands, $<$ 2 GeV and above 2 GeV. The 
0.2\,--\,2 GeV $\gamma$-rays observed by {\it Fermi}-LAT could be produced by low energy electrons through IC scattering of UV 
synchrotron photons. {\it Fermi}-LAT detected a significant variation flux in the 0.2\,--\,2 GeV band, observed over a period of 
eleven days during February 12\,--\,22, 2010. 

The observed HE ($>$2 GeV) $\gamma$-rays by {\it Fermi}-LAT and VHE $\gamma$-rays by HAGAR could be produced by IC scattering 
of the electrons having a Lorentz factor in the range $\sim 10^{4}-10^{5}$. The observations from {\it Fermi}-LAT (2\,--\,300 GeV) show a symmetrical flare centered around February 16, 2010 but VHE $\gamma$-rays flux, observed by HAGAR reaches peak with one day lag on February 17, 2010. A fresh injection of high energy electrons accelerated by shock could be responsible for the observed flare in $\gamma$-rays. 
If SSC cooling time scale is less then the light-crossing time of emission zone than the flare could be symmetric.

\subsection{Cross correlation study and time lag}

From the X-ray and $\gamma$-ray light curves (Figures 2 and 3), we can see that fluxes of Mrk421 in these bands are 
roughly correlated. We have investigated this correlation between X-ray and $\gamma$-ray bands using cross-correlation 
function (CCF) and using z-transformed discrete correlation function (ZCDF) \citep{1997ASSL..218..163A} a freely available FORTRAN 77 ZDCF 
code\footnote{http://www.weizmann.ac.il/home/tal/zdcf2.html}. Observed lag and 
correlation coefficient between RXTE\,--\,ASM, Swift\,--\,BAT and HAGAR using CCF and ZDCF algorithms are listed in Table 3 for various cases.
The time lag between X-ray and $\gamma$-ray emissions might impose constraints on emission region and also it could 
distinguish between SSC and External Compton (EC) models.

\section{Conclusions}

We found satisfactory fits for all the four states with one zone SSC model described in \S 5.1. The observed pre to post flare evolution of SED and light curve are explained by the model.
Changes in the physical conditions such as Doppler factor and magnetic field during these observations are indicated by modeling of the multiwavelength data. It 
appears that changes in these parameters are related to the activity of the source.  The change in jet flow from `State 1' to `State 2' may lead to a shock in jet. 
We also found small changes in particle energy density and magnetic field strength from `State 1' to `State 2'. 
The best fits obtained by fitting multiwavelength data to SSC model during the rising part of the flare ('State 2' and 'State 3') are found for power law index of the electron injection spectrum before break as $p1=2.2$. As p1=2.2 is an outcome of Fermi first order mechanism, it strengthens the belief that a strong shock might have accelerated the electrons in the emission zone and caused the flare by SSC mechanism. Very recently, \citep {2011ApJ...736..131A} showed that electrons present in the emission zone of Mrk421 might be accelerated by Fermi first order mechanism at the shock front.
The  best fit parameters from the modeling of SED using multiwavelength data for 'State 1' indicate an aged population of electrons with $p1=2.4$ was present in the emission zone. Light curve during the same period suggests that the entire volume of emission zone was radiating. The observed change in electron injection spectrum p1=2.3 during the decaying part of the flare might be considered to be an outcome of SSC cooling mechanism. 

The observed break in electron injection spectrum could also be explained as a cooling break, where escape time of an 
electron of Lorentz factor $\gamma_{c}$, equals the radiative cooling time. It is observed from our results that 
the cooling break changes with the state of the source. We have also found that both SED peaks move towards higher 
energy as the source flux increases, and come back to lower energies as it decreases. Spectral hardening is also 
observed at X-rays and $\gamma$-rays at the time of the flare (high state) (see Figure 5, Table 2). A departure from equipartition is also observed during this high state.
The emission process at the time of high activity is more complex than in the quiescent state. Energy dependent variability observed from {\it Fermi}-LAT has shown that the source is much more complex than anticipated.
The emission from Mrk421 also showed intra-day variability on 17 February 2010, in $\gamma$-ray bands above 0.2 GeV in 
{\it Fermi}-LAT data and $>$250 GeV from HAGAR data. We could not detect significantly this sub-hour scale variability from {\it Fermi}-LAT due to the large error bars. The detection of sub-hour variability from Fermi-LAT and HAGAR could constrain the size of the emission region.
The present results of correlation studies between X-ray and $\gamma$-ray flux variability are not conclusive enough to distinguish between the one zone SSC or EC model. However, the observed hardening of the emission spectra obtained from {\it Fermi}-LAT and RXTE\,--\,PCA data with a one day lag (see Figure \ref{fig:image3} and Table 
2), is inconsistent with the SSC model and opens up the possibility of multizone SSC or EC models. 
\section{Summary}
\begin{enumerate}[i.] 
\item HAGAR Cherenkov telescope has detected VHE $\gamma$-rays from the TeV blazar Mrk421 during 2010 February to April,
with the blazar being in a high state of activity above 250 GeV. The emission reached a peak on 
February 17, with a maximum flux of $\sim$7 Crab units, indicating a flare.
\item An energy dependent variation of flux in the HE $\gamma$-rays (0.1\,--\,300 GeV) has been observed by {\it Fermi}-LAT 
during the flare in February, 2010. 
\item Multiwavelength data indicate a variation in the spectral index during the high state.
\item A weak correlation between X-rays and VHE $\gamma$-ray has been found with a time lag of $\sim 1$ day.
\item The observed multiwavelength SEDs during February 13\,--\,19, 2010 indicate changes in the physical conditions such as magnetic field, Doppler factor and particle energies in the emission zone. The multiwavelength flare during February 16\,--\,17, 2010 and changes in physical conditions of the emission zone are explained as an effect of a passing shock in the jet.
\end{enumerate}

\begin{acknowledgements}
This work used results provided by the ASM/RXTE teams at MIT. This study also used Swift/BAT transient monitor results provided by the Swift/BAT team.
 This research has also made use of data obtained from the High Energy Astrophysics Science Archive Research Center (HEASARC), provided by NASA's Goddard Space Flight Center.
 Data from the Steward Observatory spectropolarimetric monitoring project were used. This program is supported by Fermi Guest Investigator grants NNX08AW56G and NNX09AU10G. Radio data at 15 Ghz
is used from OVRO 40 M Telescope and this Fermi blazar monitoring program is supported by NASA under award NNX08AW31G, and by the NSF under award 0808050. We are grateful to the engineering and technical staff of IIA and TIFR who have taken part in the construction of the 
HAGAR telescopes and contributed to the setting up of the front-end electronics and the data acquisition. We are 
grateful to Prof. R. Cowsik and Prof. B. V. Sreekantan for their keen interest and encouragement in the development of
the HAGAR facility. We also thank the anonymous referee for their suggestions which improved the manuscript.
\end{acknowledgements}

\bibliographystyle{aa}
\bibliography{ashu}

\end{document}